\documentclass[11pt]{article}
\usepackage{amssymb}
\usepackage{graphics}
\usepackage{epsfig}
\begin{document}
\title{  Neutral Higgs boson pair production via $\gamma\gamma$ collision in the
minimal supersymmetric standard model at linear colliders
\footnote{Supported by National Natural Science Foundation of
China.}} \vspace{3mm}
\author{{ Zhou Ya-Jin$^{2}$, Ma Wen-Gan$^{1,2}$, Hou Hong-Sheng $^{2}$, Zhang
Ren-You$^{2}$,}\\
{ Zhou Pei-Jun $^{2}$, and Sun Yan-Bin$^{2}$}\\
{\small $^{1}$ CCAST (World Laboratory), P.O.Box 8730, Beijing
100080, P.R.China}\\
{\small $^{2}$ Department of Modern Physics, University of Science and Technology}\\
{\small of China (USTC), Hefei, Anhui 230027, P.R.China}}
\date{}
\maketitle \vskip 12mm
\begin{abstract}
We investigate in detail the $\gamma\gamma$ fusion production
mechanisms of two neutral Higgs bosons ($h^0A^0$, $ H^0A^0$,
$h^0H^0$ and $H^0H^0$) within the framework of the mSUGRA-inspired
minimal supersymmetric standard model(MSSM) at an $e^+e^-$ linear
colliders, which provide a probe of the trilinear Higgs
self-couplings. We calculate the dependence of the production
rates on Higgs boson masses, the ratio of the vacuum expectation
values $\tan \beta$ and the CMS energy $\sqrt{s}$. We find that
the cross section for the $H^0H^0$ production at LC can reach
$0.2~fb$, while the cross section of $A^0H^0$ production is only
$10^{-4}\sim 10^{-3}~fb$ under our parameters.
\end{abstract}
 \vfill \eject \baselineskip=0.36in
\renewcommand{\theequation}{\arabic{section}.\arabic{equation}}
\renewcommand{\thesection}{\Roman{section}}
\newcommand{\nb}{\nonumber}
\makeatletter      
\@addtoreset{equation}{section}
\makeatother       
\section{Introduction}
\par
Many efforts have been devoted to searching for Higgs bosons and
the new physics beyond the standard model (SM) \cite{int
higgs1,int higgs2}, among which the minimal supersymmetric
standard model (MSSM) \cite{int higgs3} is the most promising one.
Five physical Higgs bosons are predicted: one CP-odd Higgs boson
($A^0$), two CP-even Higgs bosons ($h^0$ and $H^0$) and two
charged Higgs bosons ($H^+$ and $H^-$). Until now all those Higgs
bosons haven't been directly explored yet, only LEP2 group
presents the strongest lower mass limits of $91.0~GeV$ and
$91.9~GeV$ for the light CP-even and the CP-odd neutral Higgs
bosons $h^{0}$ and $A^{0}$. For a top quark mass less than or
equal to $174.3~GeV$, LEP2 group excludes the range of
$0.5<\tan\beta<2.4$\cite{Lep2} with assuming the stop quark mixing
is maximal and the conservative values for other SUSY parameters
which affect the Higgs sector,
\par
The future linear colliders will continue the work in searching
for physical Higgs bosons. There are already some detailed designs
of linear colliders, such as NLC\cite{NLC}($\sqrt{s}=500~GeV$,
integrated luminosity $220 fb^{-1}yr^{-1}$),
JLC\cite{JLC}($\sqrt{s}=500GeV$, integrated luminosity $220~
fb^{-1}yr^{-1}$), TESLA\cite{TESLA}($\sqrt{s}=500~GeV$, integrated
luminosity $340~ fb^{-1}yr^{-1}$) and CLIC\cite{CLIC} ($1~TeV <
\sqrt{s} <5~TeV$, with a luminosity of $10^{35}~ cm^{-2} s^{-1}$
at $\sqrt{s}=3~ TeV$). An $e^+e^-$ LC can also be designed to
operate as a $\gamma\gamma$ collider. This is achieved by using
Compton backscattered photons in the scattering of intense laser
photons on the initial $e^+e^-$ beams\cite{Com}. The resulting
$\gamma -\gamma$ center of mass system (CMS) energy is peaked at
about $0.8\sqrt{s}$ for the appropriate choices of machine
parameters. In $\gamma\gamma$ collision mode at the high energy
peak, we may get approximately the same luminosity as that of
$e^+e^-$ collision. Therefore, a photon LC provides additional
opportunities for hunting Higgs bosons.
\par
As we know, the trilinear Higgs boson couplings can be probed in
Higgs boson pair production processes. Detailed examinations of
Higgs boson production and decay processes are necessary in order
to detect and distinguish the signals of Higgs bosons from
background. The $Z^0h^0$, $A^0W^{\pm}$, $H^{\pm}W^{\mp}$
associated production processes at LC were studied in
Refs.\cite{Gounaris, yinjun, eehw, eehwa, eehwb, eehwc, eehwd},
which have advantages in searching for heavy Higgs bosons. The
$h^0A^0$ pair production at one-loop level via $e^+e^-$ collisions
has been discussed in Ref.\cite{Driesen}. The neutral Higgs boson
pair productions $e^+ e^- \to h^0h^0, H^0H^0, H^0h^0, A^0A^0$ and
$e^+ e^- \to Z^0h^0h^0,\nu\bar{\nu}h^0h^0$ at LC were studied by
A. Djouadi et al., \cite{Djouadi,Djouadi1}. The neutral Higgs
boson pair productions ($H^0H^0$, $H^0h^0$, $h^0h^0$, $A^0A^0$) at
the LHC were investigated in
Refs.\cite{Kniehl-a-a,9910400,Djouadi2}. The cross section for SM
neutral Higgs boson pair production in $\gamma\gamma$ collision
has been evaluated in Ref.\cite{jikia}. The $\gamma \gamma \to
h^0h^0$ and $\gamma \gamma \to A^0A^0$ pair productions in the
THDM and MSSM has been calculated respectively in
Ref.\cite{sunlazhen} and \cite{9710424, Gounaris2}.
Ref.\cite{9710424} concludes that the cross section of $e^+e^- \to
\gamma \gamma \to h^0h^0$ process in the framework of the MSSM at
LC depends on $m_{h^0}$, $\tan\beta$, photon collision modes, and
the mixing between stops. While in Ref.\cite{Gounaris2} it is
found that for an usual mSUGRA set of parameters, $\sigma (\gamma
\gamma \to A^0 A^0) \sim (0.1-0.2)\rm fb$ is expected.

\par
In this paper we study the loop induced $h^0A^0$, $ H^0A^0$,
$h^0H^0$ and $H^0H^0$ pair productions via $\gamma\gamma$
collisions at LC in the mSUGRA-inspired MSSM. We arrange the paper
as follows: In section II, we present the analytical calculation.
Numerical results and discussions are given in section III.
Section IV is a short summary.

\par
\section{Cross Section Calculation}
\par
In our calculation we use the 't Hooft-Feynman gauge. In the loop
diagram calculation we adopt the definitions of one-loop
integral functions in reference \cite{s13}. The numerical
calculation of the vector and tensor loop integral functions
are broken down to scalar loop integrals (Ref.\cite{s14}).
The Feynman diagrams and the relevant amplitudes are
generated by FeynArts package automatically \cite{denner-a-1}. The
numerical calculation of the loop integrals are implemented by
Mathematica package.

\par
\subsection{Calculation of the subprocess $\gamma\gamma \to \phi_{1}\phi_{2}$}
\par
We denote the subprocess under investigation as
\begin{equation}
\gamma(k_1)+\gamma(k_2) \rightarrow \phi_{1}(p_1)+ \phi_{2}(p_2),
\end{equation}
where $\phi_{1}\phi_{2}$=$h^0A^0$, $ H^0A^0$, $h^0H^0$ and
$H^0H^0$, $k_1$, $k_2$, $p_1$ and $p_2$ are the momenta of the
incoming photons and outgoing Higgs bosons. As the subprocess
$\gamma\gamma \rightarrow \phi_{1}\phi_{2}$ is loop-induced at the
lowest order, the one-loop order calculation can be carried out by
summing all unrenormalized reducible and irreducible one-loop
diagrams and the results will be finite and gauge invariant.

\par
We give the Feynman diagrams of the $H_iA^0$ and $H_iH^0$
production processes in Fig.\ref{feynfig1} and Fig.\ref{feynfig2}
respectively, where $H_i(i=1,2)$ represents $h^0$ and $H^0$. The
possible corresponding Feynman diagrams created by exchanging the
initial photons or the final Higgs bosons, are also involved in
our calculation. We find that the contributions from $Z^0$
exchanging s-channel Feynman diagrams with fermion loops are very
small because of the Furry theorem, so we do not show them in
Fig.\ref{feynfig1} and remove them from our calculation. We also
neglect the squark and the slepton loop diagrams in
Fig.\ref{feynfig1} because of their vanishing contributions. That
is because the contribution to the subprocesses $\gamma\gamma
\rightarrow h^0A^0,~H^0A^0$ from the diagram with a
squark(slepton) in loop which goes clockwise cancels exactly with
that which goes counterclockwise.

\par
We denoted $\theta$ as the scattering angle between one of the
photons and one of the final Higgs bosons. Then in the center of
mass system we express all the four-momenta of the initial and
final particles by means of the $\gamma\gamma$ CMS energy
$\sqrt{\hat{s}}$ and the scattering angle $\theta$. The
four-momentum components $(E,p_x,p_y,p_z)$ of final particles
$\phi_{1}$ and $\phi_{2}$ can be written as
\begin{eqnarray} \nb
\label{monmentum}\nb
p_{1}&=&\left(E_{\phi_{1}},
\sqrt{E_{\phi_{1}}^{2}-m_{\phi_{1}}^{2}}\sin\theta,~ 0,~
\sqrt{E_{\phi_{1}}^{2}-m_{\phi_{1}}^{2}}\cos\theta\right),\\
p_{2}&=&\left(E_{\phi_{2}},
-\sqrt{E_{\phi_{2}}^{2}-m_{\phi_{2}}^{2}}\sin\theta,~0,~
-\sqrt{E_{\phi_{2}}^{2}-m_{\phi_{2}}^{2}}\cos\theta\right),
\end{eqnarray}
where
$$
E_{\phi_{1}}=\frac{1}{2}\left(\sqrt{\hat{s}}+(m_{\phi_{1}}^{2}-m_{\phi_{2}}^{2})/\sqrt{\hat{s}}
\right),~~~~~
E_{\phi_{2}}=\frac{1}{2}\left(\sqrt{\hat{s}}-(m_{\phi_{1}}^{2}-m_{\phi_{2}}^{2})/\sqrt{\hat{s}}
\right).
$$
With above expressions for the four-momenta of the final state
Higgs bosons in Eq.(\ref{monmentum}), we can get that the
three-momenta of the Higgs bosons satisfy the following relation
$$
|\vec{p}_1|^2=|\vec{p}_2|^2=E_{\phi_{1}}^{2}-m_{\phi_{1}}^{2}=E_{\phi_{2}}^{2}-m_{\phi_{2}}^{2}=\frac{1}{4}
\left[\hat{s}+(m_{\phi_{1}}^{2}-m_{\phi_{2}}^{2})^2/\hat{s}-2
(m_{\phi_{1}}^{2}+m_{\phi_{2}}^{2})\right].
$$
\par
The 4-momenta of the initial photons $k_{1}$ and $k_{2}$ are
$$
k_{1}=\left(\frac{1}{2}\sqrt{\hat{s}}, 0, 0,
\frac{1}{2}\sqrt{\hat{s}}\right), ~~~~
k_{2}=\left(\frac{1}{2}\sqrt{\hat{s}}, 0, 0,
-\frac{1}{2}\sqrt{\hat{s}}\right).
$$
The Mandelstam variables are defined as
$$
\begin{array} {lll}
    \hat{s} & =(k_{1}+k_{2})^2=(p_{1}+p_{2})^2,
\end{array}
$$
$$
\begin{array} {lll}
    \hat{t} & =(k_{1}-p_{1})^2=(k_{2}-p_{2})^2,
\end{array}
$$
$$
\begin{array} {lll}
   \hat{u} & =(k_{1}-p_{2})^2=(k_{2}-p_{1})^2.
\end{array}
$$
\par
With the definitions above, we obtain the amplitude of the
subprocess $\gamma\gamma\to\phi_{1}\phi_{2}$ with simple form. For
the final states of $h^0A^0$(or $H^0A^0$), the explicit expression
of the amplitude can be written as:
\begin{eqnarray}
\label{amplitude}\nb
 {\cal M}^{H_iA^0}&=& \epsilon_{\mu}(k_1) \epsilon_{\nu}(k_2)
      \left \{f^{(i)}_1g^{\mu\nu}+f^{(i)}_{2}k_{1}^{\nu}p_{1}^{\mu}+
      f^{(i)}_{3}p_{1}^{\mu}p_{1}^{\nu}+
      f^{(i)}_{4}k_{2}^{\mu}p_{1}^{\nu}+f^{(i)}_{5}k_{2}^{\mu}p_{2}^{\nu}\right. \\ \nb
     && + \left. f^{(i)}_6k_{1}^{\nu}k_{2}^{\mu}
     + f^{(i)}_7p_{2}^{\mu}p_{2}^{\nu}+
      f^{(i)}_{8}k_{1}^{\nu}p_{2}^{\mu}+
      f^{(i)}_{9}p_{1}^{\nu}p_{2}^{\mu}+
      f^{(i)}_{10}p_{1}^{\mu}p_{2}^{\nu}+
       f^{(i)}_{11}\epsilon^{\alpha\beta\mu\nu}k_{1\alpha}k_{2\beta} \right. \\ \nb
     &&+
     \left.f^{(i)}_{12}\epsilon^{\alpha\mu\nu\beta}k_{1\alpha}p_{1\beta}+
      f^{(i)}_{13}\epsilon^{\alpha\mu\nu\beta}k_{1\alpha}p_{2\beta}+
      f^{(i)}_{14}\epsilon^{\alpha\mu\nu\beta}k_{2\alpha}p_{1\beta}+
      f^{(i)}_{15}\epsilon^{\alpha\mu\nu\beta}k_{2\alpha}p_{2\beta} \right. \\ \nb
      &&+ \left.
      f^{(i)}_{16}\epsilon^{\alpha\beta\mu\sigma}k_{1\alpha}k_{2\beta}p_{1\sigma}k_{1}^{\nu}
      +f^{(i)}_{17}\epsilon^{\alpha\beta\mu\sigma}k_{1\alpha}k_{2\beta}p_{2\sigma}k_{1}^{\nu} \right. \\
&&+\left.f^{(i)}_{18}\epsilon^{\alpha\beta\mu\sigma}k_{1\alpha}k_{2\beta}p_{1\sigma}p_{1}^{\nu}
\right \},
\end{eqnarray}
where $H_i$ represents $h^0$ or $H^0$, and
$f^{(i)}_{j}(j=1,...,18$, $i=1$ for $h^0A^0$, $i=2$ for $H^0A^0$)
are form factors. The expressions of these form factors for the
two subprocesses are very similar to each other except some
interaction vertices. We do not list their explicit expressions in
this paper because they are very lengthy.\footnote{The Mathematica
program codes of all the form factors for $\gamma\gamma \to
h^0A^0, H^0A^0$ and $\gamma\gamma \to h^0H^0,H^0H^0$ are
obtainable by sending email to zhouyj@mail.ustc.edu.cn .}
\par
For the final states of $h^0H^0$ (or $H^0H^0$), the explicit
expressions of the amplitudes can be written as:
\begin{eqnarray}
\label{amplitude2} \nb
 {\cal M}^{H_kH^0}&=&\epsilon _{\mu}(k_1) \epsilon_{\nu}(k_2)
      \left \{
       f^{(k)}_1g^{\mu\nu}+f^{(k)}_2k_{1}^{\nu}k_{2}^{\mu}+f^{(k)}_3k_{1}^{\nu}p_{1}^{\mu}
       +f^{(k)}_4k_{2}^{\mu}p_{1}^{\nu}+f^{(k)}_5p_{1}^{\mu}p_{1}^{\nu}+
       f^{(k)}_6k_{1}^{\nu}p_{2}^{\mu}
       \right. \\ \nb
&&+\left. f^{(k)}_7p_{1}^{\nu}p_{2}^{\mu}+
       f^{(k)}_8k_{2}^{\mu}p_{2}^{\nu}+f^{(k)}_9p_{1}^{\mu}p_{2}^{\nu}+
       f^{(k)}_{10}p_{2}^{\mu}p_{2}^{\nu}+
       f^{(k)}_{11}\epsilon^{\alpha\beta\mu\nu}k_{1\alpha}k_{2\beta}+
       f^{(k)}_{12}\epsilon^{\alpha\mu\nu\beta}k_{1\alpha}p_{1\beta}
       \right. \\
&&+\left.f^{(k)}_{13}\epsilon^{\alpha\beta\mu\nu}k_{2\alpha}p_{1\beta}+
       f^{(k)}_{14}\epsilon^{\alpha\beta\mu\sigma}k_{1\alpha}k_{2\beta}p_{1\sigma}k_{1}^{\nu}+
       f^{(k)}_{15}\epsilon^{\alpha\beta\mu\sigma}k_{1\alpha}k_{2\beta}p_{1\sigma}p_{1}^{\nu} \right \} ,
\end{eqnarray}
where $f^{(k)}_{j}(j=1,...15$, $k=1$ for $h^0H^0$, $k=2$ for
$H^0H^0$) are form factors.
\par
The total cross section for $\gamma\gamma \rightarrow
\phi_{1}\phi_{2}$ can be expressed in the form
\begin{equation}
\hat{\sigma}(\hat{s})=\frac{1 }{16\pi \hat{s}^2}
\int_{\hat{t^{-}}}^{\hat{t^{+}}} d\hat{t} \overline{\sum_{spin}}
\vert {\cal M}\vert^{2}
\end{equation} In the above equation,
$\hat{t}^\pm=\left[ (m^{2}_{\phi_{1}}+m^{2}_{\phi_{2}}
-\hat{s})\pm \sqrt{(m^{2}_{\phi_{1}}+m^{2}_{\phi_{2}}-\hat{s})^2
-4m^{2}_{\phi_{1}}m^{2}_{\phi_{2}}}~ \right]/2$ , and the bar over
summation means to take average over the initial polarizations.
For the subprocess $\gamma\gamma \to H^0H^0$, an additional factor
$\frac{1}{2}$ is multiplied due to the identical final particles.

\par
\subsection{Cross section of $e^{+}e^{-} \to \gamma\gamma \rightarrow \phi_{1}\phi_{2}$ process at LC}
\par
By integrating over the photon luminosity in an $e^+e^-$
linear collider, the total cross section of the process
$e^{+}e^{-} \to \gamma\gamma \rightarrow \phi_{1}\phi_{2}$ can be
obtained in the form
\begin{eqnarray}
\label{integration}
\sigma(s)= \int_{E_{min}/ \sqrt{s}} ^{x_{max}} d z \frac{d%
{\cal L}_{\gamma\gamma}}{d z} \hat{\sigma}(\gamma\gamma \to
\phi_{1}\phi_{2} \hskip 3mm
 at \hskip 3mm  \hat{s}=z^{2} s)
\end{eqnarray}
where $E_{min}=m_{\phi_{1}}+m_{\phi_{2}}$, and
$\sqrt{s}$($\sqrt{\hat{s}}$) being the
$e^{+}e^{-}$($\gamma\gamma$) CMS energy. $\frac{d\cal
L_{\gamma\gamma}}{d z}$ is the distribution function of photon
luminosity, which is defined as:
\begin{eqnarray}
\frac{d{\cal L}_{\gamma\gamma}}{dz}=2z\int_{z^2/x_{max}}^{x_{max}}
 \frac{dx}{x} F_{\gamma/e}(x)F_{\gamma/e}(z^2/x)
\end{eqnarray}
For the initial unpolarized electrons and laser photon beams, the
energy spectrum of the back scattered photon is given by
\cite{photon spectrum}
\begin{eqnarray}
\label{structure}
F_{\gamma/e}=\frac{1}{D(\xi)}\left[1-x+\frac{1}{1-x}-
\frac{4x}{\xi(1-x)}+\frac{4x^{2}}{\xi^{2}(1-x)^2}\right]
\end{eqnarray}
where
\begin{eqnarray}
D(\xi)=(1-\frac{4}{\xi}-\frac{8}{{\xi}^2})\ln{(1+\xi)}+\frac{1}{2}+
  \frac{8}{\xi}-\frac{1}{2{(1+\xi)}^2},
\end{eqnarray}
\begin{eqnarray}
  \xi=\frac{4E_0 \omega_0}{{m_e}^2}.
\end{eqnarray}
$m_{e}$ and $E_{0}$ are the mass and energy of the electron
respectively, $\omega_0$ is the laser-photon energy, and $x$
represents the fraction of the energy of the incident electron
carried by the backscattered photon. In our evaluation, we choose
$\omega_0$ such that it maximizes the backscattered photon energy
without spoiling the luminosity through $e^{+}e^{-}$ pair
creation. Then we have ${\xi}=2(1+\sqrt{2})$, $x_{max}\simeq
0.83$, and $D(\xi)=1.8397$, as used in Ref.\cite{photon para}.

\par
\section{Numerical results and discussions}
\subsection{Input parameters}
\par
In the following numerical evaluation, we present the results of
the total cross sections of the neutral Higgs boson pair
production at LC. The SM parameters are taken as : $ m_u=66$
MeV, $m_d=66$ MeV, $ m_c=1.35$ GeV, $m_s=0.15$ GeV, $ m_t=174.3$
GeV, $m_b=4.3$ GeV, $m_Z = 91.1876~GeV$, $m_{W}=80.423~GeV$ and
$\alpha=1/127.9$\cite{mass}.

\par
We take the MSSM parameters constrained in the framework of the
mSUGRA-inspired scenario as implemented in the program package
SuSpect2.1\cite{suspect}. In this scenario, only five
sypersymmetry parameters should be input, namely $M_0$,
$M_{1/2}$, $A_0$, sign of $\mu$ and $\tan\beta$, where $M_0$,
$M_{1/2}$ and $A_0$ are the scalar mass at GUT scale, the
universal gaugino mass and the trilinear soft breaking parameter
in the superpotential terms, respectively. In the package
SuSpect2.1 all the soft SUSY breaking parameters at the weak scale
are obtained through Renormalization Group Equations(RGE's)
\cite{RGE}. It uses 2-loop RGE's for the gauge, Yukawa and gaugino
couplings and all other RGE's are at 1-loop order. In this work,
we take $M_{1/2}$=130 GeV, $A_0$=200 GeV and $\mu>0$. $M_0$ is
obtained quantitatively from the $m_{A^0}$(or $m_{H^0}$) and
$\tan\beta$ values.

\vskip 10mm
\par
\subsection{Discussion and analysis}
\par
We depict the curves of the production rates of the processes
$e^+e^- \to \gamma\gamma \to \phi_{1}\phi_{2}$ in
Fig.\ref{amfig3}-\ref{fig11}. We note that if we take
$m_{A^0}=200~GeV$(or $m_{H^0}=200~GeV$ and $300~GeV$) with our
chosen input mSUGRA parameters($M_{1/2}=130~GeV$, $A_0=200~GeV$
and $\mu>0$), the mSUGRA-inspired MSSM does not permit $\tan\beta$
to be smaller than $3$ due to the experimental lower bound on the
lightest Higgs boson mass $m_{h^0}$ in MSSM ($m_{h^0}>91.0~GeV$)
\cite{Lep2}. Therefore, we take $\tan\beta$ varying in the range
of $4 \le \tan\beta \le 38$ with the above input parameters. For
the same reason, we also exclude other parameter space regions
which are inconsistent with the experimental lower bound on the
$m_{h^0}$. That is why in Fig.\ref{amfig3} and Fig.\ref{bmfig5}
the mass of Higgs boson $A^0$ varies from the value of $190~GeV$,
and the curves in Fig.\ref{cmfig7} and Fig.\ref{dmfig9} start from
the Higgs boson mass $m_{H^0}=195~GeV$(for $\tan\beta=4$) and
$m_{H^0}=200~GeV$(for $\tan\beta=30$) separately.
\par
Fig.\ref{amfig3}-\ref{btfig6} are the figures for the processes
$e^{+}e^{-} \to \gamma\gamma \to h^0A^0,~H^0A^0$ at a  LC.
Fig.\ref{amfig3} displays the cross sections of the process
$e^{+}e^{-}\to \gamma\gamma\to h^{0}A^{0}$ as the functions of the
CP-odd Higgs boson $A^{0}$ mass (or $m_{h^0}$) with $\tan \beta=4$
and $30$ respectively with $e^+e^-$ colliding energy
$\sqrt{s}=1~TeV$. The figure shows there are sophisticated
structures on both two curves, which are mainly induced by the
resonance effect. For example, at the vicinities of
$m_{A^0}=m_{\chi_2^+}+m_{\chi_1^+} \sim 300~GeV$(for
$\tan\beta=4$), $m_{A^0}=m_{\chi_1^+}+m_{\chi_2^+} \sim
290~GeV$(for $\tan\beta=30$), $m_{A^0}=2m_{\chi_2^+} \sim
570~GeV$(for $\tan\beta=4$), and $m_{A^0}=2m_{\chi_2^+} \sim
430~GeV$(for $\tan\beta=30$), we can see there are visible peaks
on the two curves due to the resonance effect from loop diagrams.

\par
We plot the total cross sections of $h^0A^0$ pair productions at
an $e^+e^-$ LC with $\sqrt{s}=1~TeV$ as the functions of $\tan
\beta$ (or $m_{h^0}$) in Fig.\ref{atfig4}. There $A^0$ mass is
taken to be $m_{A^{0}}=200~GeV$ and $300~GeV$ respectively. The
curve for $m_{A^0}=200~GeV$ shows that in the region of $\tan\beta
\le 5$, the cross section goes down rapidly with $\tan \beta$
increasing, and then goes up with $\tan \beta$ varying from $5$ to
$38$. While the curve for $m_{A^0}=300~GeV$ shows the cross
section decreases in the range of  $\tan\beta \le 5$, but goes up
smoothly when $\tan\beta \ge 5$. That is because the coupling
strengths between Higgs boson and quark pair are related with the
ratio of the vacuum expectation values.

\par
Fig.\ref{bmfig5} displays the cross sections of the process
$e^{+}e^{-} \to \gamma\gamma \to H^{0}A^{0}$ as the functions of
the $A^{0}$ mass with $\tan \beta=4$ and $30$, respectively. The
figure shows that the cross section decreases with the increment
of the mass of the CP-odd Higgs boson $A^0$. We can see that the
large $\tan\beta$ (i.e. $\tan\beta=30$) can enhance the cross
section, while the moderate value of $\tan\beta$ may suppress the
cross section.

\par
Fig.\ref{btfig6} demonstrates the cross sections of the $H^0A^0$
pair production process $e^{+}e^{-}\to \gamma\gamma\to H^{0}A^0$
at a LC with the CMS energy $\sqrt{s}=1~TeV$, as the functions of
$\tan\beta$ with $m_{A^{0}}=200$ and $300~GeV$, respectively. In
the value range of $\tan\beta < 5$, both curves for
$m_{A^0}=200~GeV$ and $300~GeV$ decrease quickly, while in the
range of $\tan \beta > 5$ they go up with the increment of
$\tan\beta$. In the region of $\tan \beta > 5$, the curve for
$m_{A^0}=200~GeV$ increases more quickly than that for
$m_{A^0}=300~GeV$ with the increment of $\tan\beta$.

\par
Fig.\ref{cmfig7}-\ref{dtfig10} are the figures for the processes
$e^{+}e^{-}\to \gamma\gamma\to h^0H^0,~H^0H^0$ at a LC.
Fig.\ref{cmfig7} shows the relationship between the cross section
of the parent process $e^{+}e^{-} \to \gamma\gamma \to h^0H^0$ and
$m_{H^0}$ with $\sqrt{s}=1~TeV$. The large suppression on the
curve of $\tan\beta=4$ around the position of $m_{H^0}
=m_{\chi_1^+}+m_{\chi_2^+} \sim 300 GeV$, corresponds to the
resonance effect. Again due to the resonance effect, the curve of
$\tan\beta=30$ shows a large suppression at
$m_{H^0}=m_{\tilde{\tau}_1}+m_{\tilde{\tau}_2} \sim 205~GeV$ and a
large enhancement at $m_{H^0}=2 m_t \sim 350~GeV$. We can see from
the figure that the cross section with $\tan\beta=4$ is larger
than the corresponding one with $\tan\beta=30$ in most of the
parameter region.
\par
In Fig.\ref{ctfig8}, the curves show that the cross sections of
the process $e^{+}e^{-}\to \gamma\gamma\to h^0H^0$ decrease with
the increment of $\tan\beta$ in the region of $\tan\beta<10$ (for
$m_{H^0}=200~GeV$) and $\tan\beta<15$ (for $m_{H^0}=300~GeV$), and
then their values become to be not sensitive to $\tan\beta$.
\par
Fig.\ref{dmfig9} indicates the dependence of the cross section of
$H^0H^0$ production process at LC on the mass of $H^0$ with
$\sqrt{s}=1~TeV$. We find that the cross section decreases quickly
with the increment of the mass of $H^0$. Fig.\ref{dtfig10} shows
the dependence of the cross section of $e^{+}e^{-}\to
\gamma\gamma\to H^0H^0$ on $\tan\beta$ with $m_{A^0}=200~GeV$ and
$300~GeV$ respectively. Both curves show that the cross sections
go down at the low $\tan\beta$ region with the increment of $\tan
\beta$, and it changes smoothly in the moderate $\tan \beta$
range. When $\tan\beta > 25$, the curve for $m_{A^0}=300~GeV$ goes
up slowly while the curve for $m_{A^0}=200~GeV$ keeps unchanged.
The cross section of $H^0H^0$ pair production at $e^+e^-$ LC is in
the order of $10^{-1}$ to $10^{-2}$ in our chosen parameter space,
and specially when $m_{H^0}=195~GeV$ and $\tan\beta=4$, the cross
section reaches $0.13 fb$.

\par
We also plot the total cross sections of the $h^0A^0$, $H^0A^0$,
$h^0H^0$ and $H^0H^0$ production processes as the functions of
$\sqrt{s}$ at LC in Fig.\ref{fig11}. There we take the input
parameters $M_{1/2}=130~GeV$, $A_0=200~GeV$, $\mu>0$,
$\tan\beta=4$ and $m_{H^0}=200~GeV$. These input parameters induce
$m_{h^0}=92.31~GeV$ and $m_{A^0}=193.6~GeV$ in the mSUGRA-inspired
MSSM. From this figure we find that the cross section for $H^0H^0$
production at LC is larger than the cross sections of other three
channels of Higgs boson pair production processes. The cross
section $\sigma(e^{+}e^{-} \to \gamma\gamma \to H^0H^0)$ is of the
order $10^{-2} \sim 10^{-1}$, and can exceeds $0.2~fb$ in some
parameter space. It shows that this process could be visible at
linear colliders with high enough luminosities. The curve for the
cross section of $e^+e^- \to \gamma\gamma \to H^0A^0$ has the
lowest value among the four curves in our chosen parameter space,
which has the values in the order of $10^{-4} \sim 10^{-3}~fb$.

\section{ Summary}
\par
In this paper, We evaluate the loop induced production processes
of two neutral Higgs bosons via $\gamma\gamma$ fusion, i.e.
$e^+e^- \to \gamma\gamma \rightarrow \phi_{1}\phi_{2}$
($\phi_{1}\phi_{2}=h^0A^0,H^0A^0,h^0H^0$ and $H^0H^0$), within the
framework of the mSUGRA-inspired minimal supersymmetric standard
model at $e^+e^-$ linear collider. We analyze the dependence of
the total cross sections on Higgs boson masses
$m_{\phi_{i}}({\phi_{i}}=H^0,A^0)$, the ratio of the vacuum
expectation values $\tan \beta$ and colliding energy $\sqrt{s}$,
respectively. With the same input parameters in Higgs sector(i.e.
$\tan\beta=4$, $m_{H^0}=200~GeV$), we find that the $H^0H^0$-pair
production at LC ($e^+e^-\to\gamma\gamma \to H^0H^0$) has larger
production rate than the other three channels of the neutral Higgs
boson pair productions in general, and can reach $0.2~fb$
quantitatively in our chosen parameter space, which could be
visible for LC with high enough luminosity. While the cross
sections of $e^+e^- \to \gamma\gamma \rightarrow
h^0A^0,H^0A^0,h^0H^0$ at LC have smaller cross sections. The
$H^0H^0$-pair production has the lowest production rate among all
the four channels in our chosen parameter space, which has the
values of the order $10^{-4} \sim 10^{-3}~fb$.

\par
\noindent{\large\bf Acknowledgments:} This work was supported in
part by the National Natural Science Foundation of China and a
grant from the University of Science and Technology of China.

\begin{figure}[htp] 
\centering \epsfig{file=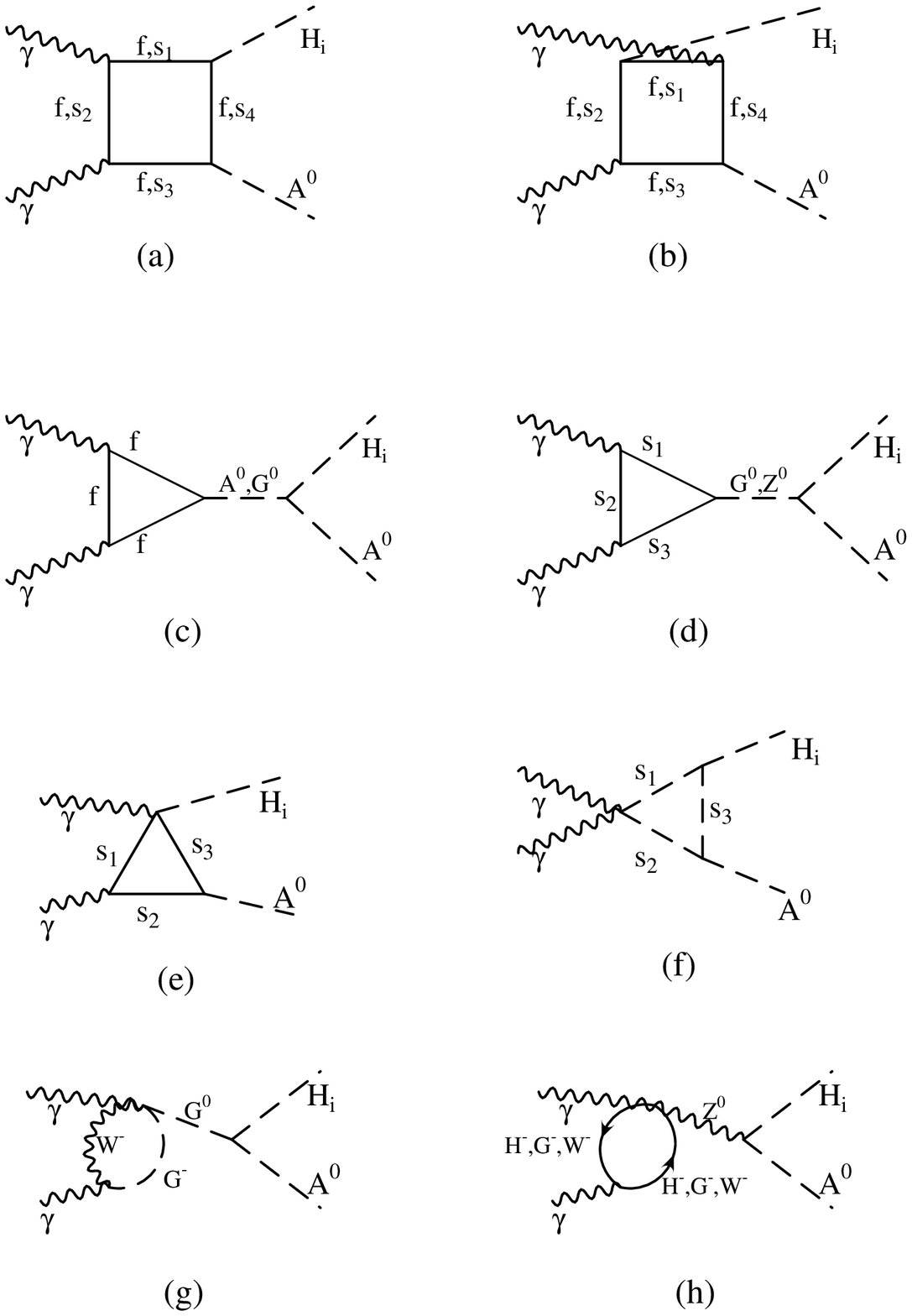, height=7.2 in, width=5.6 in}
\end{figure}

\begin{figure}[htp] 
\centering \epsfig{file=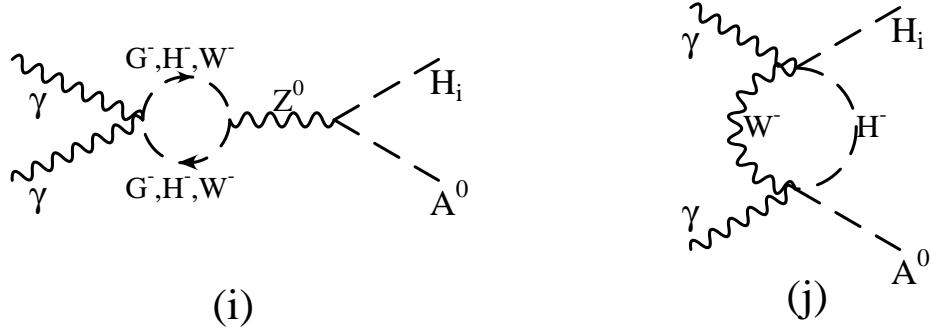, height=1.8 in, width=5 in}
\caption{\label{feynfig1} {
 The Feynman diagrams of the subprocess $\gamma \gamma \to H_{i}A^0$,
where $H_{i}$ denotes $h^0$ and $H^0$, $f$ can be quark, lepton or
chargino, and $\tilde{f}$ is squark or slepton. The meaning of
other notations in the figures are as follows: (a):
$s_1s_2s_3s_4$={\scriptsize$G^-G^-G^-H^-$,$G^-G^-W^-H^-$,
$W^-G^-G^-H^-$, $W^-G^-W^-H^-$, $G^-W^-G^-H^-$, $G^-W^-W^-H^-$,
$W^-W^-G^-H^-$, $W^-W^-W^-H^-$, $H^-H^-H^-G^-$, $H^-H^-H^-W^-$};
(b):$s_1s_2s_3s_4$={\scriptsize$G^-H^-H^-G^-$, $G^-H^-H^-W^-$,
$W^-H^-H^-G^-$, $W^-H^-H^-W^-$, $H^-G^-G^-H^-$, $H^-G^-W^-H^-$,
$H^-W^-G^-H^-$, $H^-W^-W^-H^-$}; (d):when the propagator is
{\footnotesize$G^0$}, $s_1s_2s_3$=$u_{\pm}u_{\pm}u_{\pm}$,
{\scriptsize $G^-G^-W^-$, $W^-W^-G^-$}, and when the propagator is
{\footnotesize$Z^0$}, $s_1s_2s_3$=$u_{\pm}u_{\pm}u_{\pm}$,
{\scriptsize $H^-H^-H^-$, $G^-G^-G^-$, $W^-W^-W^-$, $G^-G^-W^-$,
$G^-W^-G^-$,$W^-G^-W^-$,$W^-W^-G^-$}; (e):
$s_1s_2s_3$={\scriptsize $W^-G^-H^-$, $H^-H^-W^-$, $W^-W^-H^-$};
(f): $s_1s_2s_3$={\scriptsize $H^-H^-G^-$, $G^-G^-H^-$,
$W^-W^-H^-$, $H^-H^-W^-$.}}}
\end{figure}

\begin{figure}[htp] 
\centering \epsfig{file=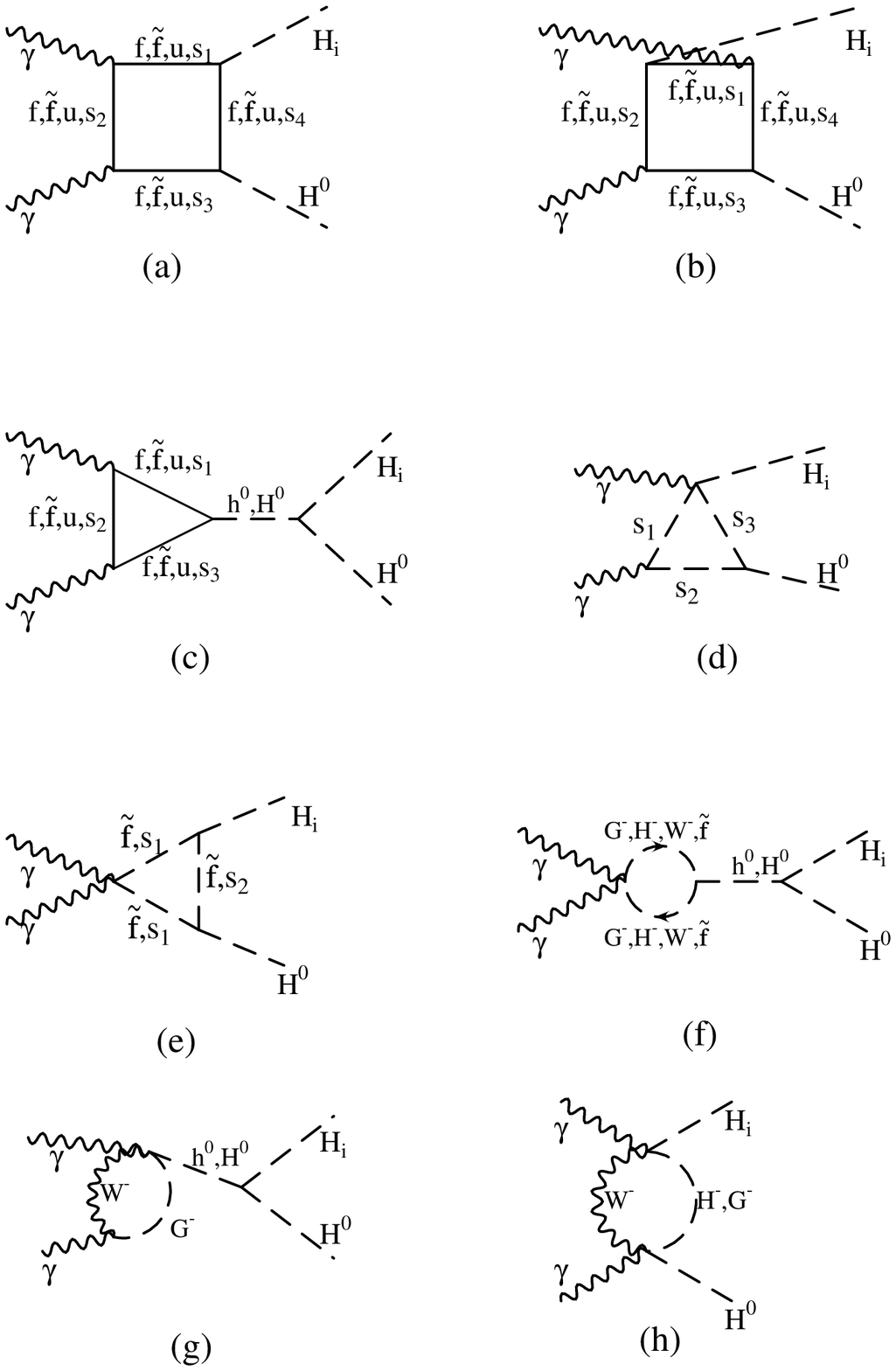, height=7.2 in, width=5.6in}
\end{figure}

\begin{figure}[htp] 
\centering \epsfig{file=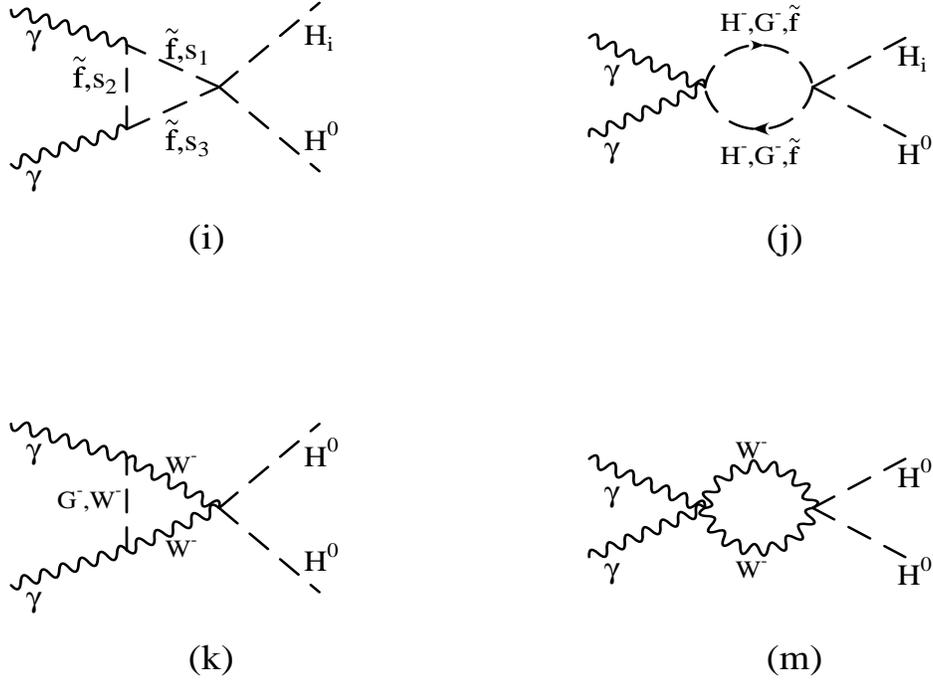, height=3.6 in, width=5 in}
\caption{\label{feynfig2} {
 The Feynman diagrams of the subprocess $\gamma \gamma \to$ {\small $H_{i}H^0$},
 where {\small $H_{i}$} denotes $h^0$ and {\small $H^0$}, $f$ can be quark, lepton or chargino,
 $\tilde{f}$ is squark or slepton, and $u$ denotes $u_+$ or $u_-$(the ghost field of {\small $W$} boson).
 The figures form (a) to (j) are for both the
 {\small $h^0H^0$} and {\small $H^0H^0$} production processes,
while (k) and (m) are for the {\small $H^0H^0$} final states only.
$s_i$($i=1,...4$) can be {\footnotesize$H^-$,$G^-$} or
{\footnotesize $W^-$}, and they appear in the figures in the
following way: (a):there are two cases permitted:
 {\scriptsize(i)}$s_1=s_2=s_3$; {\scriptsize(ii)}
 $s_1s_2s_3=${\scriptsize $G^-W^-G^-$, $G^-W^-W^-$, $G^-G^-W^-$, $W^-W^-G^-$, $W^-G^-G^-$,
 $W^-G^-W^-$}.
(b):there are five cases permitted:
{\scriptsize}$s_1=s_2=s_3=s_4$;
 {\scriptsize(ii)}$s_1=s_2\ne s_3=s_4$;
 {\scriptsize(iii)}$s_1=s_4$, and $s_2s_3=${\scriptsize $G^-W^-$ or $W^-G^-$},
 {\scriptsize(iv)}$s_2=s_3$,
 and $s_1s_4=${\scriptsize $G^-W^-$ or $W^-G^-$}. {\scriptsize(v)}$s_1s_4=${\scriptsize $G^-W^-$ or
 $W^-G^-$},
 and $s_2s_3=${\scriptsize $G^-W^-$ or $W^-G^-$}.
 (c):$s_1s_2s_3=$, {\scriptsize$H^-H^-H^-$, $G^-G^-G^-$,
 $W^-W^-W^-$, $G^-G^-W^-$, $G^-W^-G^-$, $W^-G^-W^-$, $G^-W^-W^-$}.
 (d): $s_1s_2s_3=${\scriptsize$H^-H^-W^-$, $G^-G^-W^-$,
 $W^-G^-H^-$, $W^-G^-G^-$, $G^-W^-W^-$, $W^-W^-H^-$, $W^-W^-G^-$}.
 (i): $s_1s_2s_3=$ {\scriptsize $H^-H^-H^-$, $G^-G^-G^-$,
 $G^-W^-G^-$}.
}}
\end{figure}

\begin{figure}
\includegraphics*[25pt,25pt][560pt,400pt]{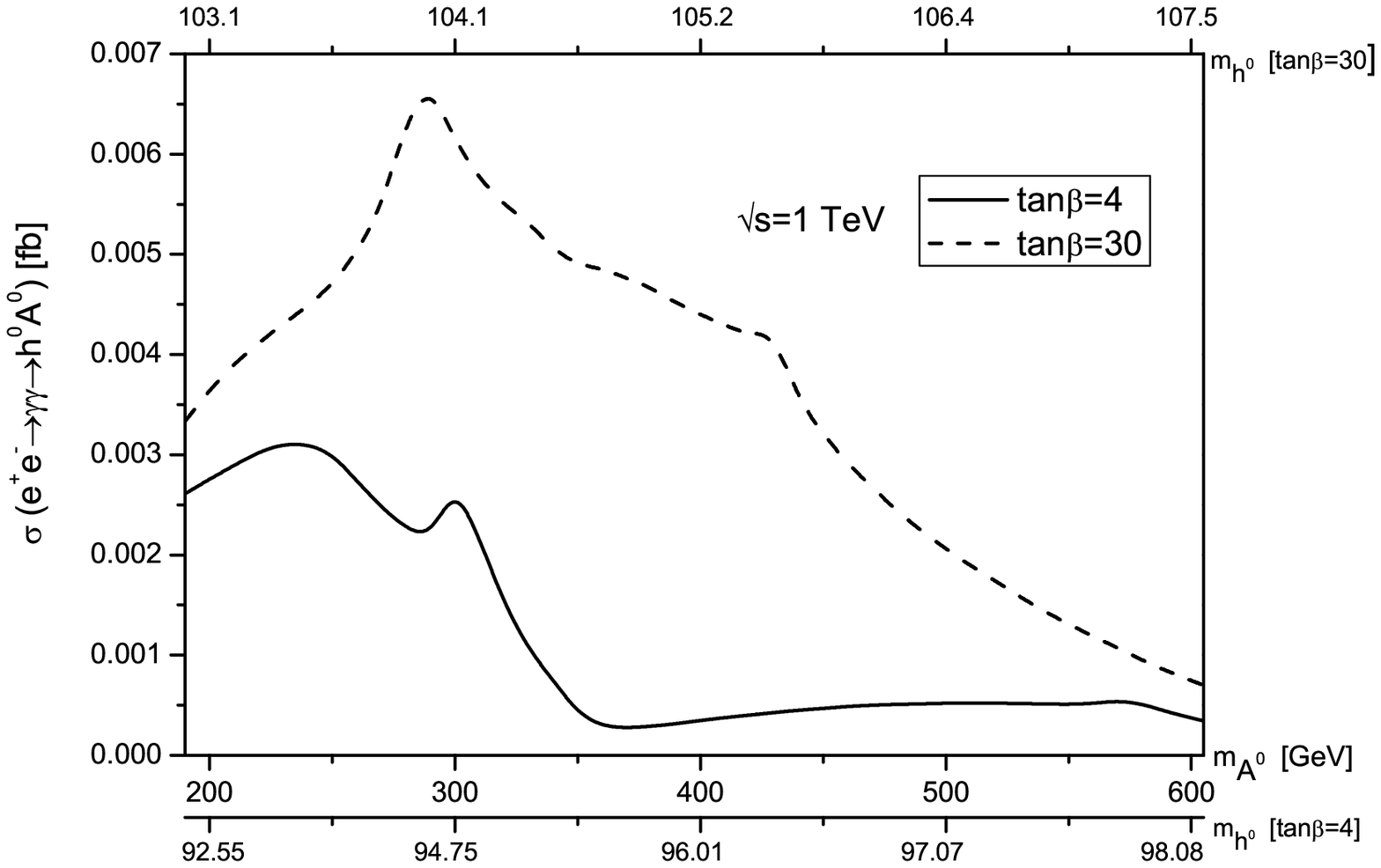}
\caption{\label{amfig3}The cross sections $\sigma$ of the process
$e^+e^-\to \gamma\gamma \to h^0A^0$, as the functions of the mass
of Higgs boson $A^0$(starting at $m_{A^0}=190~GeV$), and $\tan
\beta$ is taken as $4$ and $30$, respectively.}
\end{figure}

\begin{figure}
\includegraphics*[25pt,25pt][560pt,400pt]{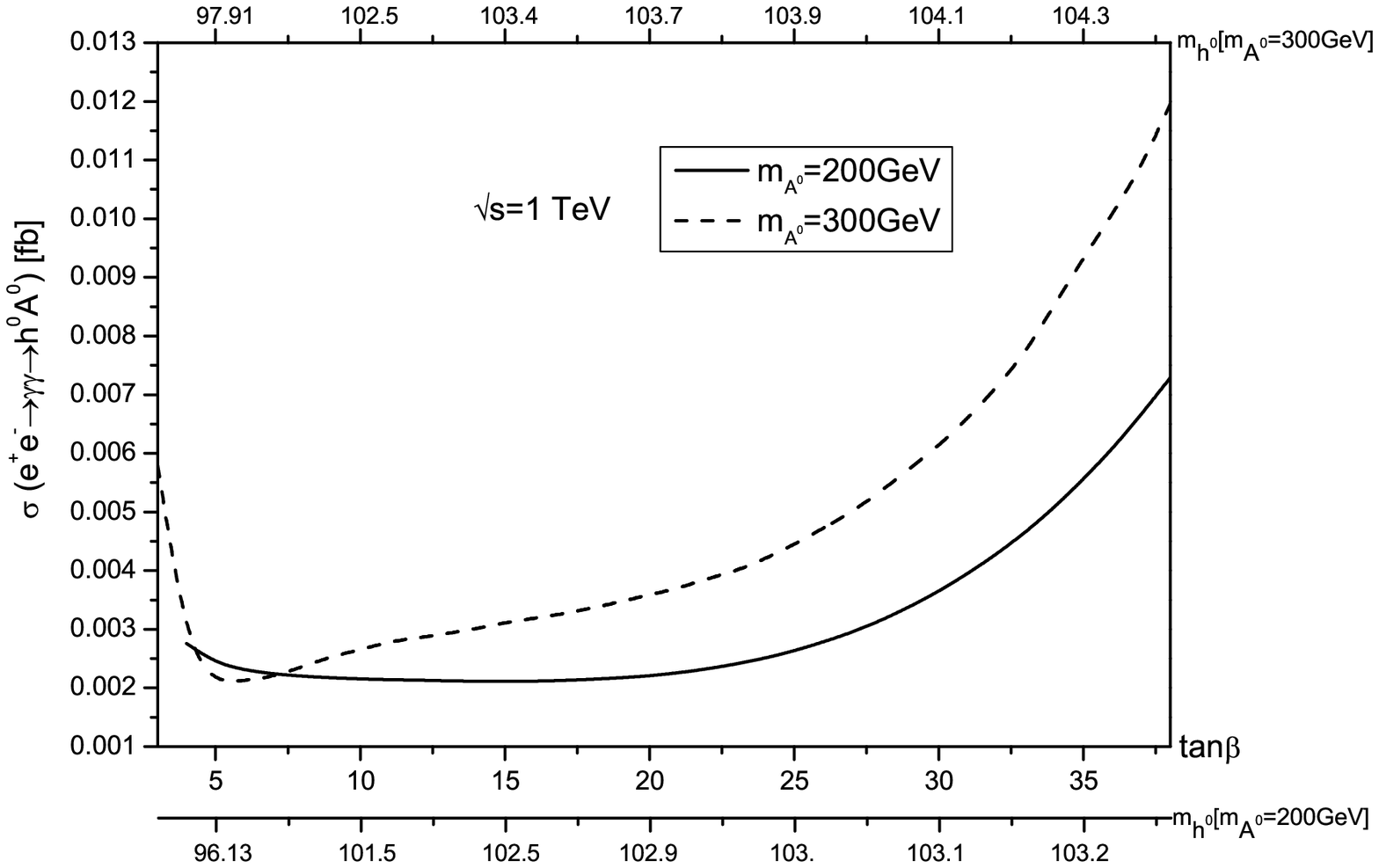}
\caption{\label{atfig4}The cross sections $\sigma$ of the process
$e^+e^-\to \gamma\gamma \to h^0A^0$, as the functions of $\tan
\beta$. The mass of Higgs boson $A^0$ is taken as
$200~GeV$(starting at $\tan\beta=4$) and $300~GeV$(starting at
$\tan\beta=3$), respectively.}
\end{figure}

\begin{figure}
\includegraphics*[25pt,25pt][560pt,400pt]{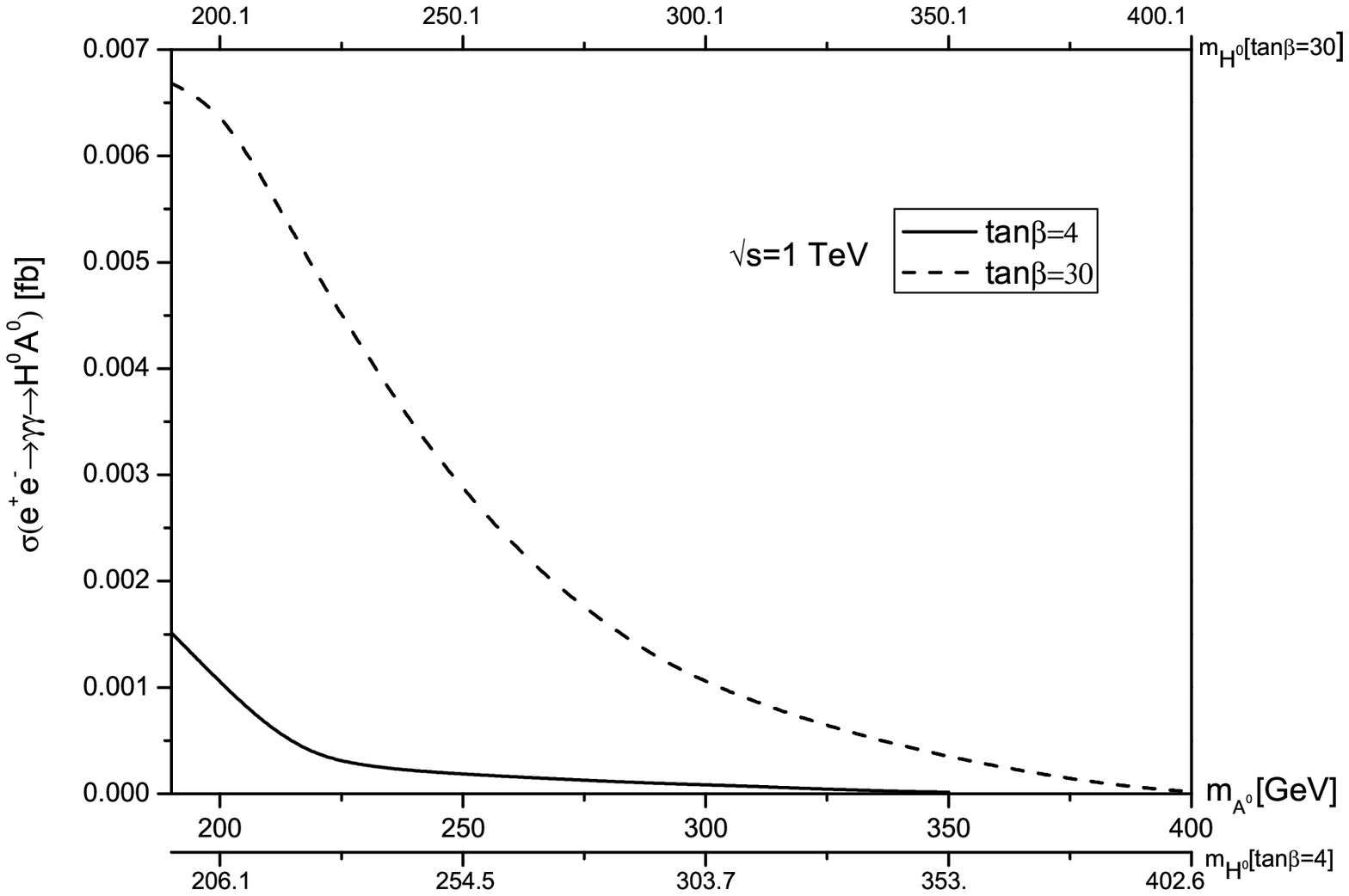}
\caption{\label{bmfig5}The cross sections $\sigma$ of the process
$e^+e^-\to \gamma\gamma \to H^0A^0$, as the functions of the mass
of Higgs boson $A^0$(starting at $m_{A^0}=190$ GeV), and $\tan
\beta$ is taken as $4$ and $30$, respectively.}
\end{figure}

\begin{figure}
\includegraphics*[25pt,25pt][560pt,400pt]{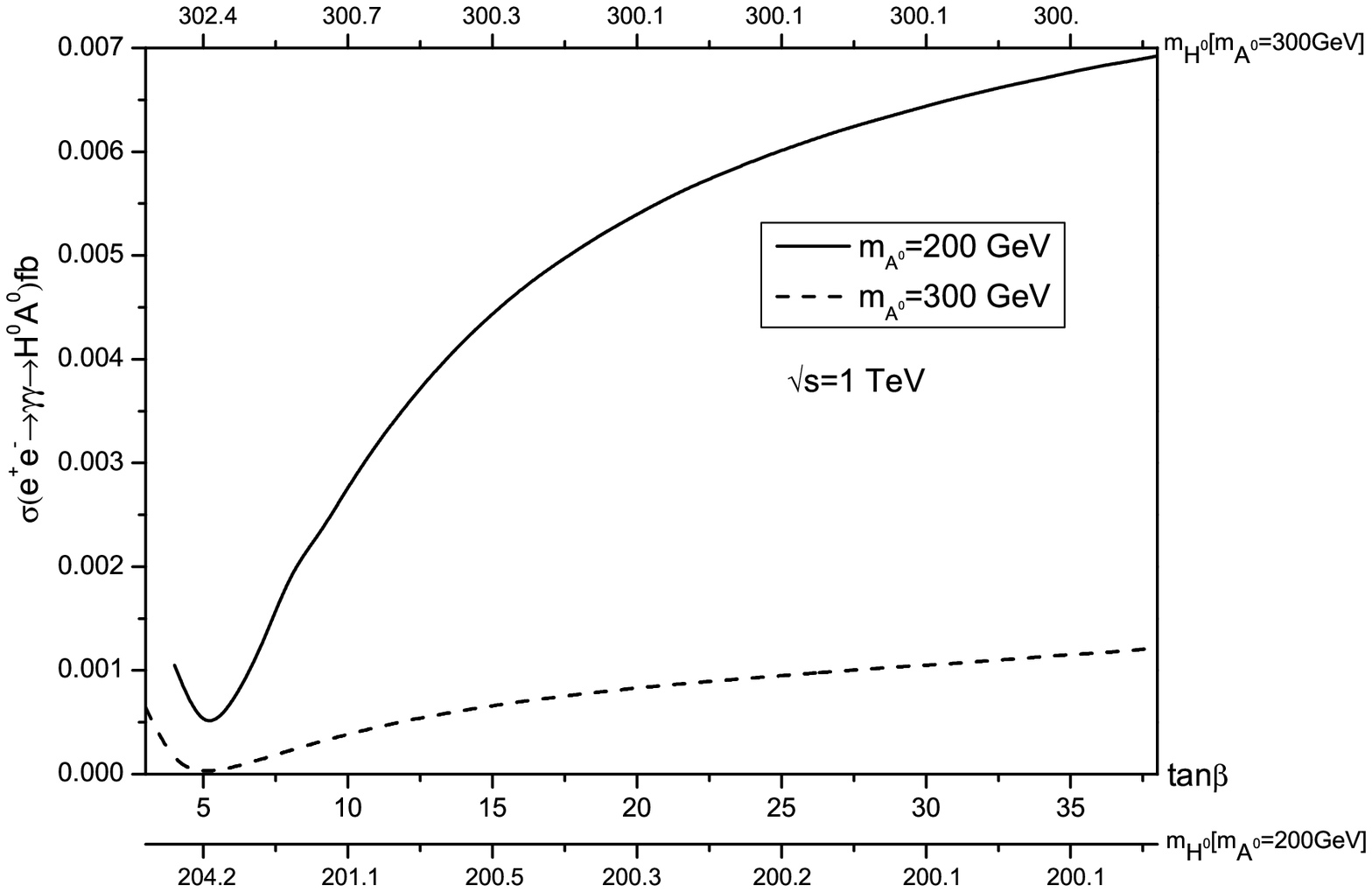}
\caption{\label{btfig6}The cross sections $\sigma$ of the process
$e^+e^-\to \gamma\gamma \to H^0A^0$, as the functions of $\tan
\beta$. The mass of Higgs boson $A^0$ is taken as
$200~GeV$(starting at $\tan\beta=4$) and $300~GeV$(starting at
$\tan\beta=3$), respectively.}
\end{figure}

\begin{figure}
\includegraphics*[25pt,25pt][560pt,400pt]{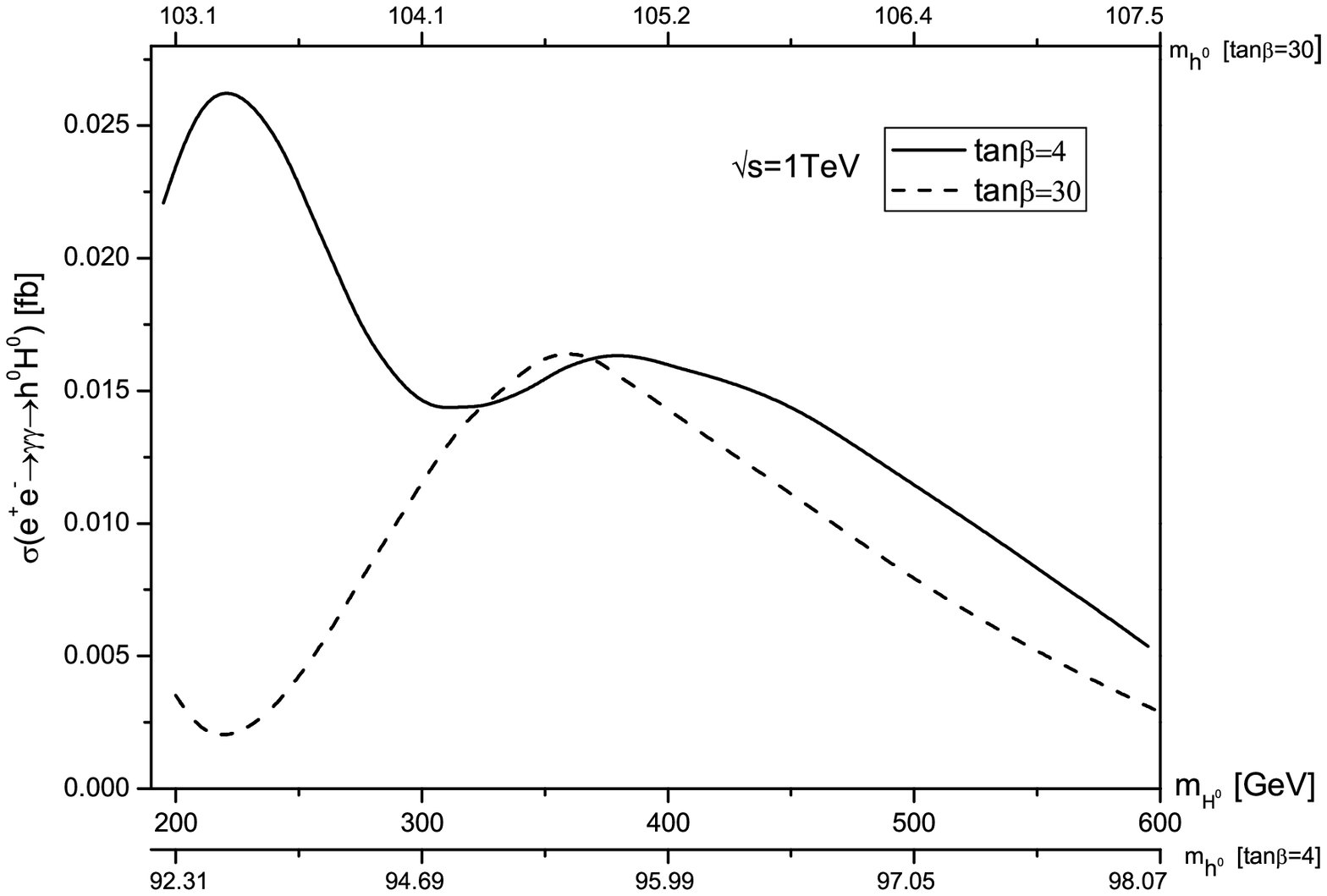}
\caption{\label{cmfig7}The cross sections $\sigma$ of the process
$e^+e^-\to \gamma\gamma \to h^0H^0$, as the functions of the mass
of Higgs boson $H^0$, and $\tan \beta$ is taken as $4$(starting at
$m_{H^0}=195~GeV$) and $30$(starting at $m_{H^0}=200~GeV$),
respectively.}
\end{figure}

\begin{figure}
\includegraphics*[25pt,25pt][560pt,400pt]{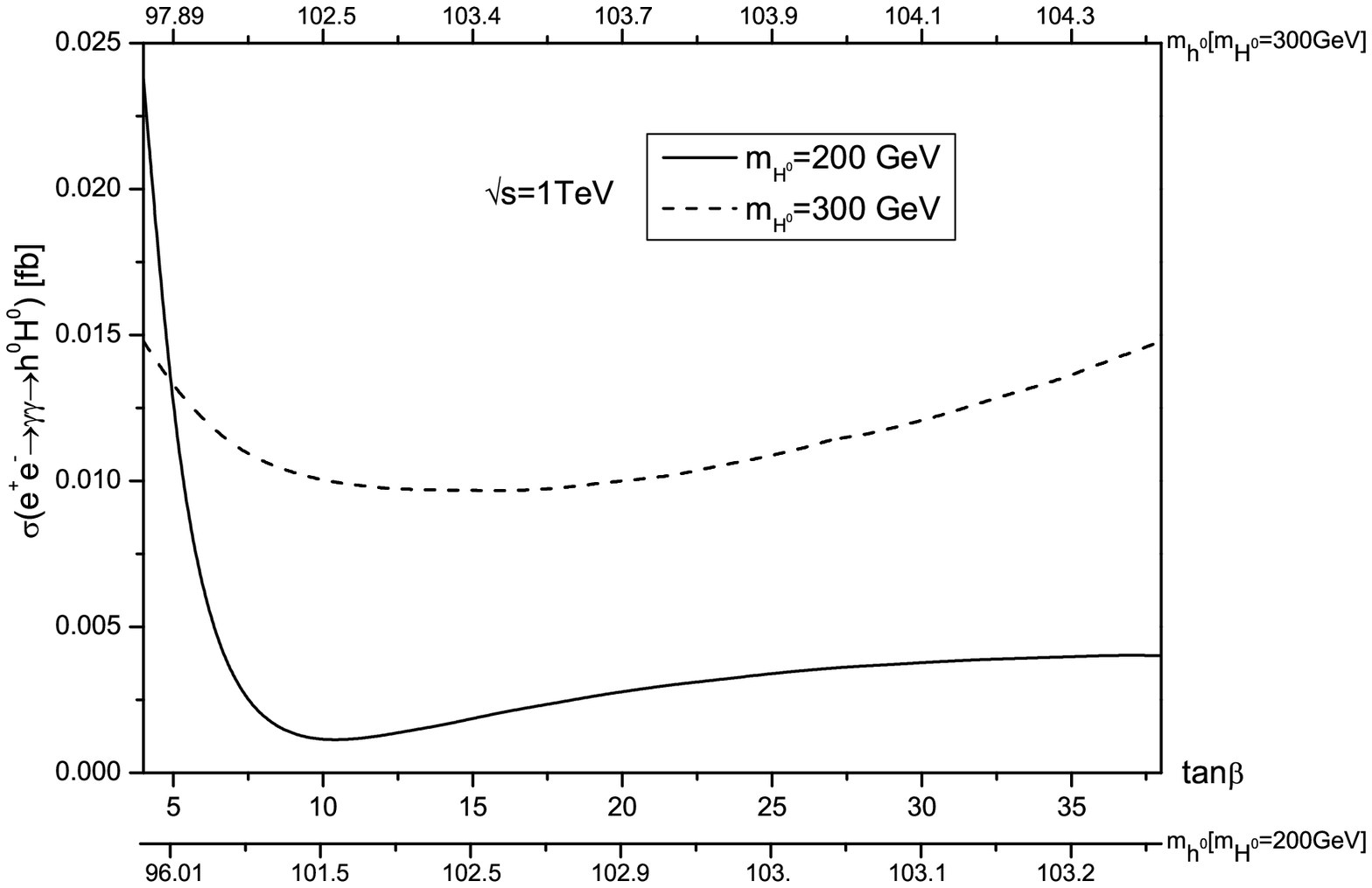}
\caption{\label{ctfig8}The cross sections $\sigma$ of the process
$e^+e^-\to \gamma\gamma \to h^0H^0$, as the functions of $\tan
\beta$(starting at $\tan\beta=4$). The mass of Higgs boson $A^0$
is taken as $200~GeV$ and $300~GeV$, respectively.}
\end{figure}

\begin{figure}
\includegraphics*[25pt,25pt][560pt,400pt]{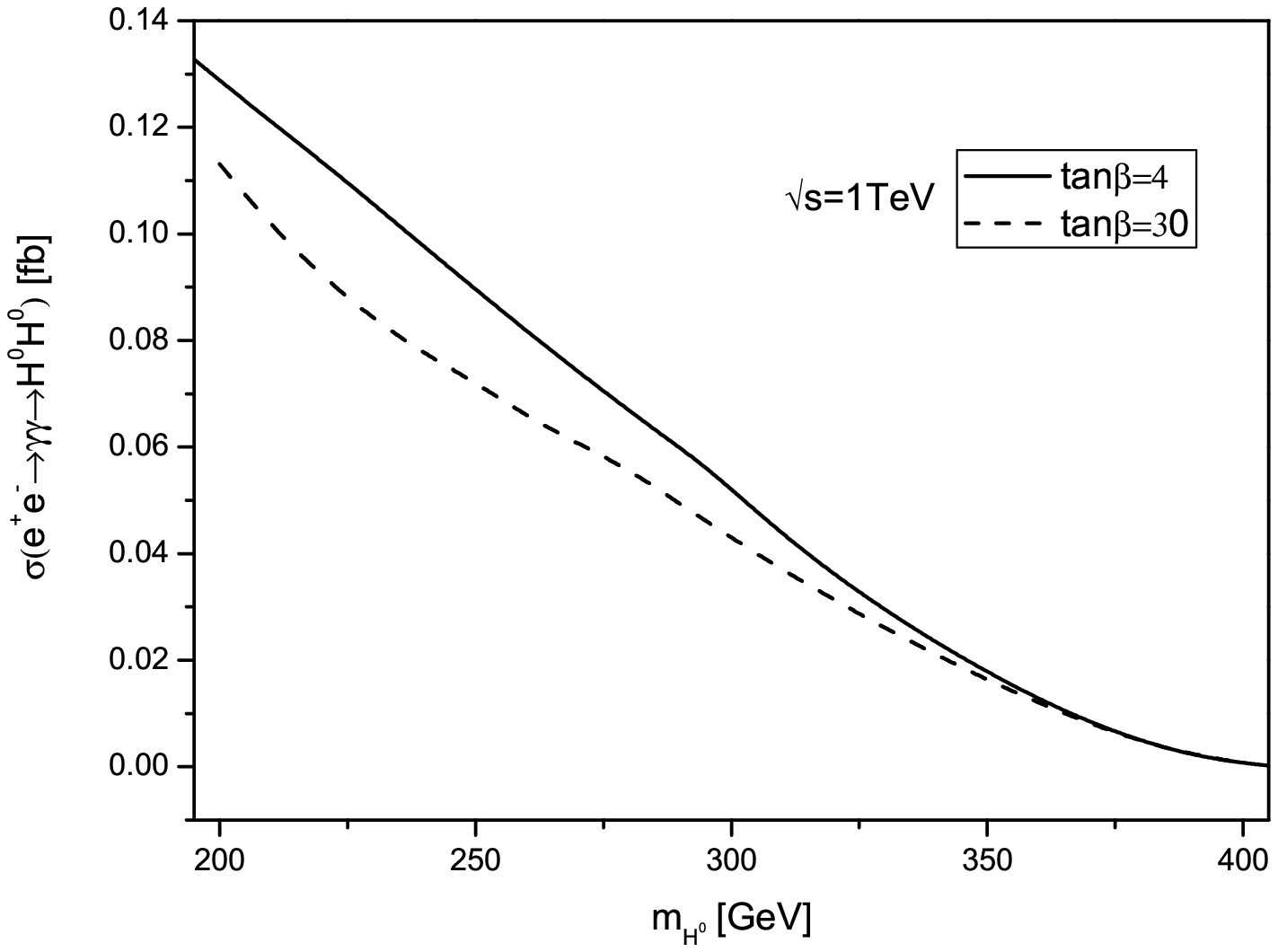}
\caption{\label{dmfig9}The cross sections $\sigma$ of the process
$e^+e^-\to \gamma\gamma \to H^0H^0$, as the functions of the mass
of Higgs boson $H^0$, and $\tan \beta$ is taken as 4(starting at
$m_{H^0}=195$ GeV) and 30(starting at $m_{H^0}=200$ GeV),
respectively.}
\end{figure}

\begin{figure}
\includegraphics*[25pt,25pt][560pt,400pt]{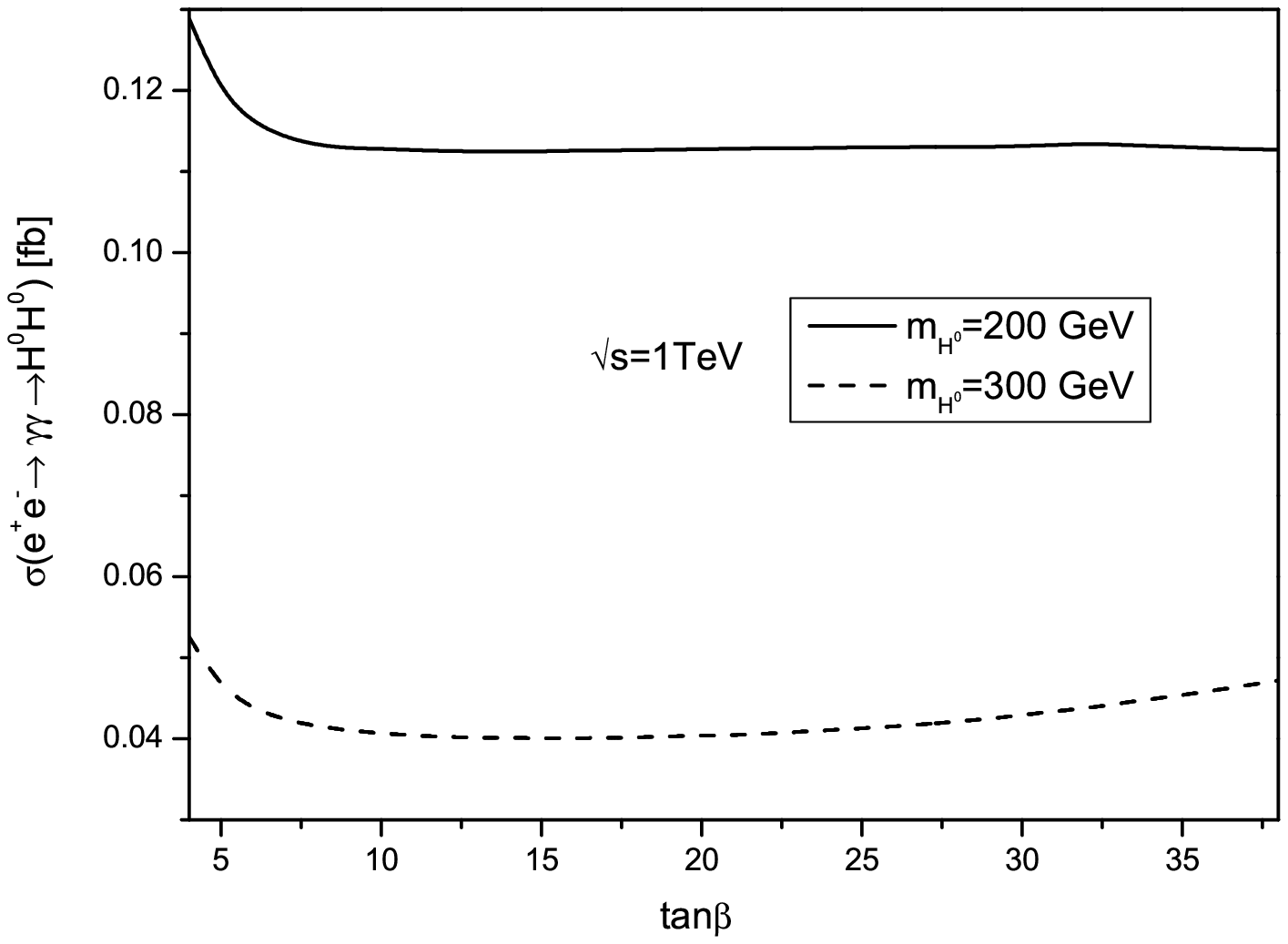}
\caption{\label{dtfig10}The cross sections $\sigma$ of the process
$e^+e^-\to \gamma\gamma \to H^0H^0$, as the functions of $\tan
\beta$(starting at $\tan\beta=4$). The mass of Higgs boson $H^0$
is taken as $200~GeV$ and $300~GeV$, respectively.}
\end{figure}

\begin{figure}
\includegraphics*[25pt,20pt][560pt,400pt]{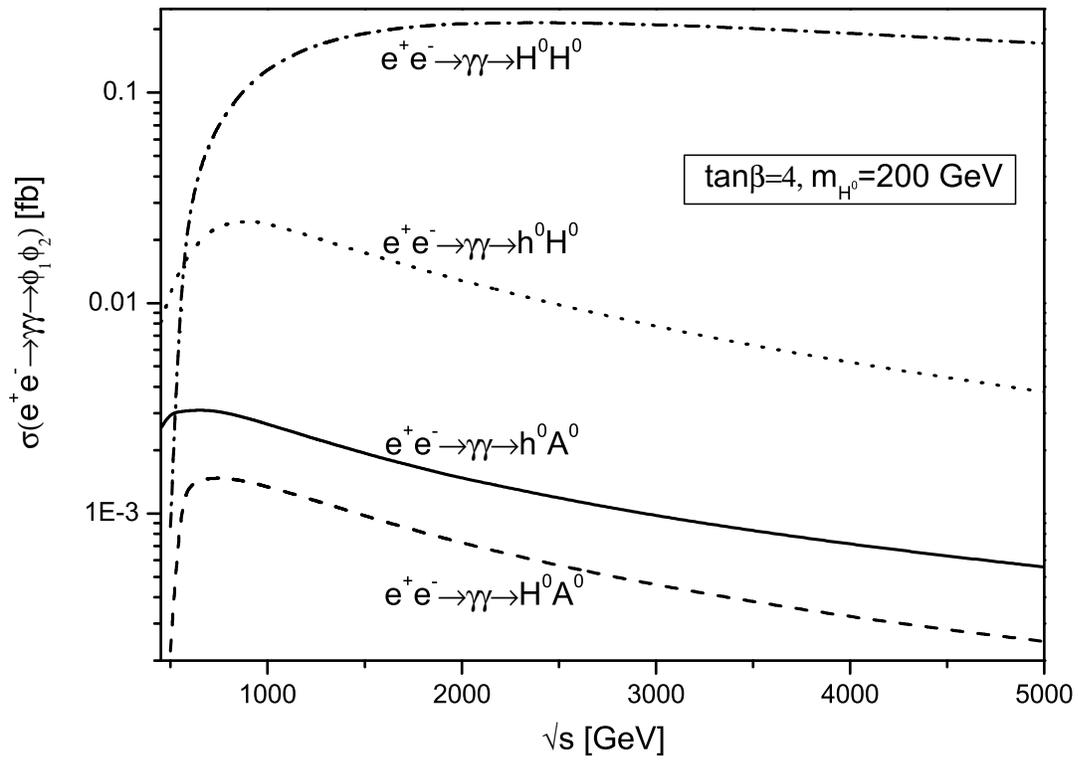}
\caption{\label{fig11}The cross sections $\sigma$ of the processes
$e^+e^-\to \gamma\gamma \to \phi_1\phi_2$, as the functions of the
$e^+e^-$ CMS energy $\sqrt{s}$.}
\end{figure}

\end{document}